\documentclass[aps,prb,twocolumn]{revtex4-1}

\usepackage{amsmath}    
\usepackage{graphicx}   
\usepackage{verbatim}   
\usepackage{color}      
\usepackage{subfigure}  
\usepackage{hyperref}   
\usepackage{color}
\usepackage{bbm}
\usepackage{caption}

\newcommand{\Tp}{\hat{T}_p}
\newcommand{\Ti}{\hat{T}_i}
\newcommand{\SG}{S_{\langle G \rangle}}
\renewcommand{\Re}{\textrm{ Re}}
\renewcommand{\Im}{\textrm{ Im}}

\begin{document}

\title{Resonant scattering induced thermopower\\ in one-dimensional disordered systems}

\author{D. M\"uller}
\email{danmuell@itp.phys.ethz.ch}
\affiliation{Institute for Theoretical Physics, ETH Zurich}

\author{W. J. Smit}
\affiliation{Institute for Theoretical Physics, ETH Zurich}

\author{M. Sigrist}
\affiliation{Institute for Theoretical Physics, ETH Zurich} 

\date{\today}

\begin{abstract}
This study analyzes thermoelectric properties of a one-dimensional random conductor which shows localization effects and simultaneously includes resonant scatterers yielding sharp conductance resonances. These sharp features give rise to a distinct behavior of the Seebeck coefficient in finite systems and incorporate the degree of localization as a means to enhance thermoelectric performance, in principle. 
The model for non-interacting electrons is discussed within the Landauer-B\"uttiker formalism such that analytical treatment is possible for a wide range of properties, if a special averaging scheme is applied. The approximations in the averaging procedure are tested with numerical evaluations showing good qualitative agreement, with some limited quantitative disagreement. The validity of low-temperature Mott's formula is determined and a good approximation is developed for the intermediate temperature range. In both regimes the intricate interplay between Anderson localization due to disorder and conductance resonances of the disorder potential is analyzed. 
\end{abstract}

\pacs{}

\maketitle

\section{Introduction}

In the context of energy harvesting, research on thermoelectrics has acquired renewed momentum, as efficient thermoelectric materials may provide a means to convert heat into electrical energy at relatively low maintenance and reliably as no engines with moving parts are involved \cite{riffat_thermoelectrics_2003,matsubara_development_2002,disalvo_thermoelectric_1999,sootsman_new_2009}. 
Thermoelectricity constitutes a standard part of transport theory: electricity and heat are connected, for instance, by the Seebeck coefficient $ S $ which relates a temperature difference $ \Delta T $ with a voltage difference $ \Delta V $ as $  \Delta V = S \Delta T $ assuming open circuit conditions \cite{Ziman-Book-1972,Smith-Jensen-Book-1989,MacDonald-2006,Zlatic-2014}.
An important quantity for applications is
the figure of merit $ Z $, a measure for the efficiency of the energy conversion for a material acting as a thermoelectric device. The dimensionless parameter $ZT$ is defined as $ ZT = \sigma_{el} S^2 T  / \kappa $, where $ \sigma_{el} $ ($ \kappa $) denote the electrical (heat) conductivity. Much effort is devoted to enhancing $ZT$ for which values beyond 1 are only rarely reported \cite{riffat_thermoelectrics_2003,matsubara_development_2002,disalvo_thermoelectric_1999}. 
Strategies for improvements have turned to nanostructuring of materials which reduces the phonon heat conductivity and increases $\sigma$ and $S$ by adjusting internal properties as mobile charge carriers confined in a narrow energy range \cite{snyder_complex_2008, chen_recent_2003, sootsman_new_2009,fogelstrom_impurity_2007,heremans_enhancement_2008}. Alternatively, also correlated systems\cite{koshibae_thermopower_2000,terasaki_novel_2010,sagarna_electronic_2012} and low dimensionality have been considered as a way to strongly suppress the phonon heat conductivity and enhance the electrical conductivity as well as the thermopower\cite{hicks_effects_1993,hicks_experimentalstudy_1996,venkatasubramanian_thin-film_2001,hsu_cubic_2004,harman_quantumdot_2002,dresselhaus_new_2007,chen_recent_2003,chen_phonon_2001,
CRC_handbook_thermo_nano_2005,zlatic_quantdots_2010,zlatic_largethermo_2011}.

In the low-temperature limit the Seebeck coefficient is given by Mott's formula, 
\begin{equation}
S = -\frac{\pi^2}{3} \frac{k_B^2 T}{e}  \left.\frac{\partial \log \sigma_{el} }{\partial E}\right|_{E=\mu} \;
\label{eq:Mott_formula}
\end{equation}
which relates $S$ to the energy dependence of the electrical conductivity $ \sigma_{el} $ at the chemical potential\cite{Ziman-Book-1972,Smith-Jensen-Book-1989,MacDonald-2006}. This formula indicates that $S$ is also a measure of the energy dependence of the conductivity.  
This aspect has been, for example, been emphasized by Mahan and Sofo who suggested to optimize the thermoelectric performance by using devices where the conductance has delta-peak like structures in the vicinity of the chemical potential \cite{mahan_best_1996}. Such kind of structures are naturally obtained in systems with conductance resonances. 

In our study we focus on disordered one-dimensional systems with disorder where we investigate the effect of Anderson localization and, through special design of our model, also the situation of mobile charge carriers near localization. Such a system can be realized by a model of randomly positioned impurities, barriers of given specifications. In general, all carriers are localized in such a system. However, if these scatterers are identical, they can develop conductance resonances at specific values of energy. In combination with the Anderson localization effect such resonances show interesting features which, in principle, could be used to design improved thermoelectric devices. Note that Anderson localization in general has been investigated by many groups for a variety of reason \cite{kapitulnik_thermoelectric_1992,enderby_electron_1994,durczewski_enhanced_1998,villagonzalo_thermoelectric_1999,
 pollak_localization_1985}.
Using the transfer matrix formulation by Landauer and B\"uttiker we will analyze different regimes of the system analytically as well as numerically. This is possible if we ignore the interaction among the electrons. 
While transport properties of electronic states near the mobility edge of systems displaying Anderson localization physics are often effectively modeled, for instance by a variable range hopping model, we have not to resort here to any effective model. Rather we will benefit from the fact that the transfer matrix approach enables us to deal with various aspects of a disordered one-dimensional system analytically. While we are mainly interested in the basic behavior of thermopower in our special type of random one-dimensional model, we will also briefly address statistical features of a finite random system for comparison with related discussions\cite{van_langen_thermopower_1998}.

\section{Model}

We first introduce a model of a disordered one-dimensional system of non-interacting electrons. Configuring the model as a random array of well localized elastic scatterers will allow us to deal with the effect of Anderson localization to a large extent analytically by using the transfer matrix approach. 

\subsection{Transfer Matrix Method}

The Landauer-B\"uttiker formalism of transfer matrices is undoubtedly the most successful method to discuss transport properties of one-dimensional mesoscopic systems, if only elastic scattering is involved which conserves the electron energy \cite{landauer_electrical_1970,buttiker_generalized_1985,sivan_multichannel_1986,guttman_thermopower_1995}. Inelastic scattering processes, such as electron-electron or electron-phonon scattering, are omitted for simplicity.

\begin{figure}[h]
\center
\includegraphics[scale=0.9]{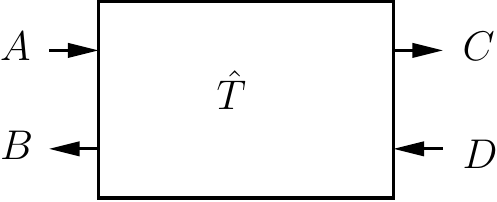} 
\caption{Notation of the in- and out-going wave functions for a given spatial range of the potential described by $\hat{T}$.}
\label{fig:transfer}
\end{figure}

The effect of scattering of an electron at a potential limited to certain range in space is encoded in the so-called transfer matrix $\hat{T}$,
\begin{equation}
\hat{T} = \begin{pmatrix}
 \frac{1}{t^*} & -\frac{r^*}{t^*} \\
 -\frac{r}{t}  & \frac{1}{t}
\end{pmatrix} \; .
\label{eq:trans_matrix}
\end{equation}
This matrix relates the in and out-coming wave functions at the left hand side, $A$ and $B$, with the out and incoming wave functions at the right hand side of the given spatial range of the potential, $C$ and $D$, see Fig.~\ref{fig:transfer}, in the following way
\begin{equation}
 \begin{pmatrix} C \\ D \end{pmatrix} = \hat{T} \begin{pmatrix} A \\ B \end{pmatrix}.
\end{equation}

The system resistance introduced by backscattering of electrons is given by
\begin{equation}
R_{sys} = R_0 \frac{|r|^2}{|t|^2}  \; ,
\end{equation}
where $R_0 = h/2e^2$ and $ 1 = |r|^2+|t|^2 $.
The measured resistance includes also the contact resistance $R_c = h/2e^2=R_0$ which gives then
\begin{equation}
 R = R_{sys} + R_c = \frac{R_0}{|t|^2}.
\end{equation}
The full conductance is the inverse of $R$,
\begin{equation}
G = G_0 |t|^2 \; ,
\label{eq:conductance}
\end{equation}
where $G_0 = 2e^2/h$ is the conductance quantum.

\begin{figure}[h]
\center
\includegraphics[scale=0.9]{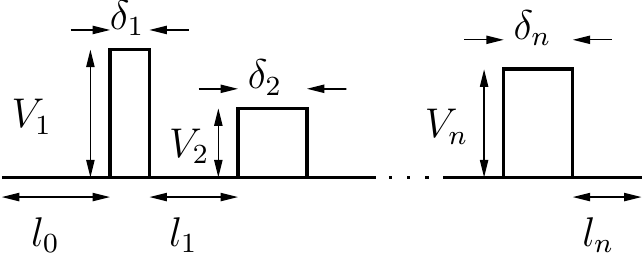} 
\caption{Sketch of the potential landscape with its underlying parameters.}
\label{fig:landscape}
\end{figure}

We model our system by well localized potential barriers. For mathematical simplicity we assume box-shaped potentials
as sketched in Fig.~\ref{fig:landscape}, which can be easily parametrized by the height $V_i$ and the width $\delta_i$ and the distance $l_i$ to the potential on the left hand side.

For a system with $ N $ barriers with a given set of parameters $\{ (l_i,V_i,\delta_i) \}_{i=1, ... N} $ we can compute the corresponding transfer matrix $\hat{T}$ by splitting it into elementary transfer matrices, $\Tp$ for the propagation between the barriers (scattering free) and $\Ti$ for the scattering part of a potential barrier,
\begin{equation}
 \hat{T} = \Tp(l_{N}) \cdot \Ti(V_N,\delta_N) \cdot \Tp(l_{N-1}) \dots  \Ti(V_1,\delta_1) \cdot \Tp(l_{0}) \; ,
 \label{eq:cond_decomp}
\end{equation}
which corresponds to a simple sequential product of transfer matrices. 
The propagation transfer matrix specifies only the change of the phase between barriers and is given by
\begin{equation}
 \Tp(l) = \begin{pmatrix} e^{i k l} & 0 \\ 0 & e^{-i k l} \end{pmatrix} \; ,
 \label{eq:transfermatrix_propagation}
\end{equation}
whereas the impurity transfer matrix is more complex. It can be computed by solving the one dimensional Schr\"odinger equation and imposing the continuity conditions at the interfaces. For our box potentials this yields after some straightforward calculation,
\begin{equation}
 \Ti(V,\delta) = 
 \begin{pmatrix}  \cos(q \delta) + \frac{i \epsilon}{2} \sin(q \delta) & \frac{i \eta}{2} \sin(q \delta) \\
                  - \frac{i \eta}{2} \sin(q \delta)    & \cos(q \delta) - \frac{i \epsilon}{2}\sin(q \delta) 
 \end{pmatrix},
 \label{eq:transfermatrix_impurity}
\end{equation}
where $k = \sqrt{2m E}/\hbar$, $q = \sqrt{2m(E-V)}/\hbar$ and
\begin{equation}
 \epsilon = \frac{q}{k} + \frac{k}{q}, \quad  \ \eta = \frac{q}{k} - \frac{k}{q}.
\end{equation}

By combining Eq.~(\ref{eq:trans_matrix},\ref{eq:conductance},\ref{eq:cond_decomp}) we can compute $G$ for any configuration of barriers in a system.
The link to thermoelectricity is obtained through the Seebeck coefficient $S$ given by the Cutler-Mott formula,
\begin{equation}
 S_G(T,\mu) = - \frac{1}{e T} \frac{\int dE \ (E-\mu) G(E) \left(-\frac{\partial f}{\partial E}\right) }{\int dE \ G(E) \left(-\frac{\partial f}{\partial E}\right)} \; ,
 \label{eq:cutler_mott}
\end{equation}
where $f$ denotes the Fermi-Dirac distribution function.

This expression for the thermopower includes the contact resistance and differs from the thermopower of the wire only. However, it has been shown \cite{guttman_thermoelectric_1995} that this discrepancy vanishes in the limit of infinite scattering centers. Therefore, we do not specifically distinguish between these two kinds of Seebeck coefficients and use the formula in Eq.~(\ref{eq:cutler_mott}).

\subsection{Averaging Procedure}
\label{sec:averag_procedure}

The Seebeck coefficient $S_G$ depends on the configuration $\{ (l_i, V_i, \delta_i) \}_i$, which constitutes a too large number of parameters. Therefore we turn here to an averaging over many configurations, assuming a certain self-averaging for large enough systems. The averaged Seebeck coefficient $\langle S \rangle$ is defined by
\begin{equation}
 \langle S \rangle = \int \ \left( \prod_i dl_i dV_i d\delta_i  \right)  P\left(\{(l_i,V_i,\delta_i)\}_i \right)  S_G \; ,
\end{equation}
with the probabilistic parameter distribution function $P$.
This definition is very general and a direct analytical evaluation of $\langle S \rangle$ is challenging. 

\begin{figure}[h]
\center
\includegraphics[scale=0.8]{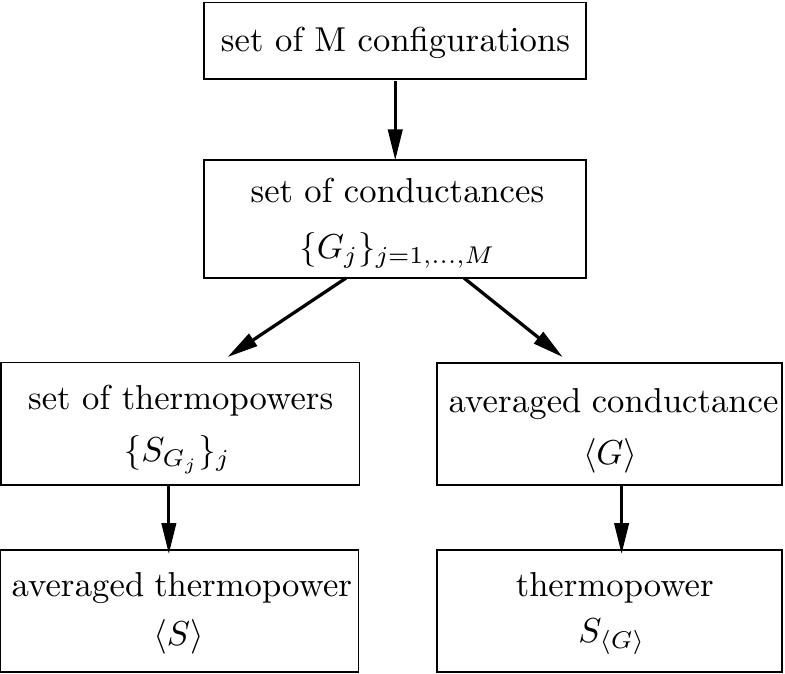}
\caption{Schematic diagram of the averaging process.}
\label{fig:average_schematic}
\end{figure}

We will work with an alternative form of averaged thermopower. The basic idea is to average first the conductance $G$ instead of $S$, see Fig.~\ref{fig:average_schematic}, and secondly, to calculate the thermopower by applying Eq.~(\ref{eq:cutler_mott}) on the average of $G$. Later we will discuss this scheme by comparing the two averaging procedure using numerics. 

We distinguish here two averages the harmonic and the arithmetic. The harmonic average of the conductance, 
$\langle G \rangle_h$, is given by
\begin{equation}
\frac{1}{\langle G \rangle_h} \equiv 
\int \left( \prod_j dl_j dV_j d\delta_j \right) P(\{(l_i,V_i,\delta_i)\}_i) \frac{1}{G} \; .
\label{eq:t_average}
\end{equation}
while the arithmetic one, $\langle G \rangle_a$, has the form,
\begin{equation}
\langle G \rangle_a \equiv 
\int \left( \prod_j dl_j dV_j d\delta_j \right) P(\{(l_i,V_i,\delta_i)\}_i) \; G \; .
\end{equation}
It turns out that the former is more easily accessible in an analytic approach and we will for the time being focus on this approach. However, later we will compare the two averages in a numerical discussion.

\subsection{Distribution of scattering barriers}

The energy dependence of the conductance is an interference effect. Whenever the width of a barrier or the distance between two potential barriers matches a multiple of the electrons wave length, constructive interference and perfect transmission through this part of the system occurs. Due to the random distribution of the scatterers such resonances are washed out. Here, however, we would like to introduce a certain distribution which allows to see perfect transmission. For this purpose we assume that the different parameters are distributed independently. Thus, we can assume the product form,
\begin{eqnarray}
 P(\{(l_i,\delta_i, V_i)\}_i) &=& \prod_j P_{i}(\delta_j,V_j) \; P_{p}(l_j)\nonumber \\
 &=& \prod_j P_l(l_j) \ P_\delta(\delta_j) \ P_V(V_j) \; ,
 \label{eq:distribution}
\end{eqnarray}
where
\begin{eqnarray}
 P_l(l_i) &=& \textrm{const} \; , \phantom{\frac{\delta}{\delta}} \\
 P_V(V_i) &=& \delta(V_0 - V_i) \; , \phantom{\frac{\delta}{\delta}} \\
 P_\delta(\delta_i) &=& \frac{1}{\sqrt{2\pi}\sigma} \exp\left( - \frac{(\delta_i-\delta)^2}{2\sigma^2}  \right) \; ,
\end{eqnarray}
using the $\delta$-function $\delta(\dots)$ in $P_V$ (not to be confused with the impurity width $\delta$).
Inter-barrier resonances are completely wiped out by this distribution and the barrier height of all scatterers is fixed. 
This will lead generally to a localization and a vanishing conductance for infinitely large system. 
However, our model has built-in the possibility for a recovery of the conductance by perfect transmission, if all barriers satisfy the resonance condition simultaneously. This can be reached for $ \sigma =0 $ where we find conductance peaks for specific resonance energies of the
electrons. These perfect conductance resonances are, however, reduced and washed out with growing standard deviation $ \sigma $. 

Thus, perfect transmission can now only be reached within the barriers and be wielded using the standard deviation~$\sigma$ of the widths~$\delta_i$.

\section{Analysis of the Seebeck coefficient}
\label{sec:results}

The above approximations allow us to continue our analytical study of thermoelectricity and will
give us good insight in the basic properties in various regimes and the important parameters. 

\subsection{Computation of $\SG$}

\subsubsection{Averaged Conductance $\langle G \rangle$}

We start with the averaging of the conductance and use, as announced above, the harmonic average which gives us a simple
analytical result. The harmonic average of $G$ defined in Eq.~(\ref{eq:t_average}) is given by 
\begin{equation}
\langle G \rangle = \frac{2 G_0}{1+\left[ \left( \hat{F}_p \hat{F}_i \right)^N\right]_{11}} \; ,
\end{equation}

\noindent where $\hat{F}_{p}$ and $\hat{F}_{i}$ are $3\times 3$ matrices which incorporates both, the distribution functions $P_{p,i}$ and the transfer matrices $\hat{T}_{p,i}$. The barrier heights $V_i$ are already fix to be $ V_0 $ according to the distribution function $ P_V$. The parameter $N$ is the number of barriers (impurities) in the system. For the explicit derivation of this result as well as the definition of  $\hat{F}_{p,i}$ we refer to Appendix \ref{app:averaging_process:result}.

The explicit form of the conductance reads, see Appendix \ref{app:harmonic_average:result}, 
\vskip0.4in
\begin{widetext}
\begin{equation}
\langle G \rangle(x) = \frac{2 G_0}{1+\left( 1+ \frac{\sin^2 \sqrt{\lambda(x-1)}}{2x(x-1)} + 
 \left( 1- e^{-2\lambda (x-1) \sigma^2/\delta^2} \right) \frac{\cos\left( 2 \sqrt{\lambda (x-1)} \right)}{4x(x-1)} 
 \right)^N } = \frac{2 G_0}{1+ | t_1(x,\lambda, \sigma)|^{-2N}}
 \label{eq:aver_trans_perfect}
\end{equation}
\end{widetext}
with the dimensionless energy ${x=E/V_0}$, the barrier parameter ${\lambda = 2 m V_0 \delta^2/\hbar^2}$ and $ | t_1 |^2 $ the
mean transmission probability for a single barrier.

\begin{figure}
\center
\includegraphics[scale=1.0]{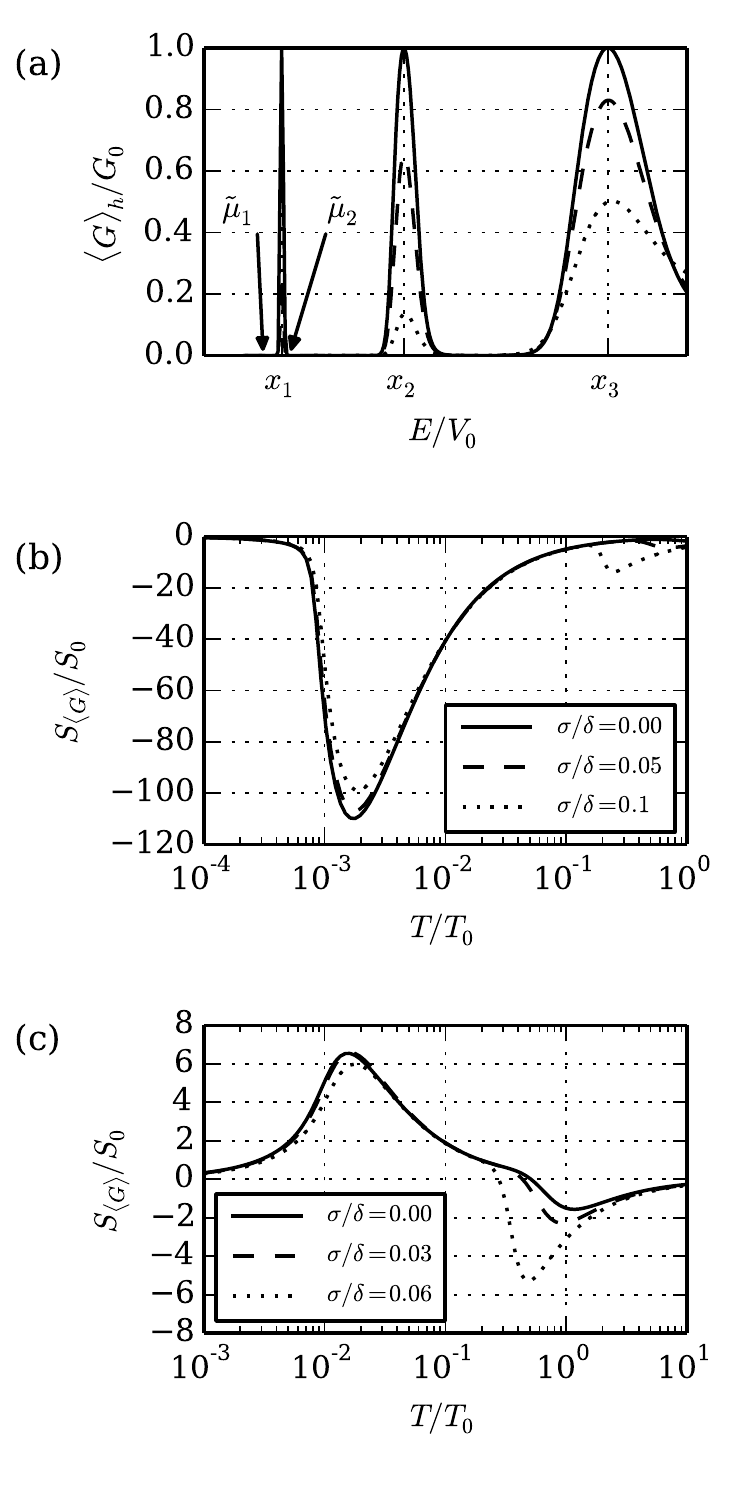}
\caption{(a) The harmonically averaged conductance for $\sigma/\delta = 0,0.03,0.06$ (solid, dashed and dotted line) at $N=1000$ and $\lambda = 9$. Based on this conductance we computed the thermopower $\SG$ for (b) $\tilde{\mu}_1 = \mu_1/V_0=1.6$ (which is below the resonance energy $E_1 = 1+\pi^2/9 \approx 2.1$) as well as (c) $\tilde{\mu}_2=2.3$ (which is above the resonance energy $E_1$) where $S_0 = k_B/e$ and $T_0 = V_0 / k_B$.}
\label{fig:conduct_seebeck}
\end{figure}

The behavior of $ \langle G \rangle $ is displayed in Fig.~\ref{fig:conduct_seebeck}(a) for various values of $ \sigma/\delta $. As is obvious from Eq.~(\ref{eq:aver_trans_perfect}) resonances appear at energy values $ x_n=1+(n \pi)^2/\lambda$ with the integer $ n \geq 1 $, yielding perfect conductance for $ \sigma =0 $ (absolutely identical barriers, randomly located), i.e. $ \langle G \rangle (x_n) = G_0 $. With increasing $ \sigma $ the conductance shrinks. Sufficiently far from the resonance energies the conductance vanishes, manifesting Anderson localization. Note that in these intermediate energy ranges the conductance can be safely approximated by
\begin{equation}
\langle G \rangle \approx 2 G_0 \left[ \left( \hat{F}_p \hat{F}_i \right)^{N}\right]_{11}^{-1} =  2 G_0| t_1(x,\lambda, \sigma)|^{2N} \; 
\label{eq:g-approx}
\end{equation}
for $ N \gg 1 $, which allows us to define a localization length $ \ell $,
\begin{equation}
\langle G \rangle = 2 G_0 e^{-L / \ell} \; ,
\end{equation}
where $ L $ is the system length with $ L n_i = N $ ($n_i $ is the constant impurity density) and $ \ell = -  (2 n_i \ln |t_1|)^{-1} $.

The conductance resonances allow us now to approach an energy regime where the Anderson localization length can become comparable to the system size and the carriers become delocalized. Additionally, the regions around the resonances are interesting for thermoelectricity due to the strong energy dependence of the conductance, as the Mott formula suggests.

\subsubsection{The Thermopower $\SG$}
\label{sect:s-aver}

In Fig.~\ref{fig:conduct_seebeck}(b,c) we show the Seebeck coefficient $\SG$ as a function of temperature in two regimes which display characteristic behaviors. The data are given in units of $S_0 = k_B/e \approx 86.2 \ \mu V/K$ and $T_0 = V_0/k_B$. 

We use $\lambda=9$ and $N=1000$ as the system parameters and check the temperature range $10^{-4}<T/T_0<10^{-1}$.

In Fig.~\ref{fig:conduct_seebeck}(b) we set the chemical potential slightly below the lowest conductance peak which yields a negative Seebeck coefficient. Under this condition the resonance closest to the chemical potential dominates the behavior of the thermoelectric effect. Using a logarithmic temperature scale we see a pronounced peak in $S_{\langle G \rangle}$. Its magnitude and the location are almost unaffected by $\sigma$. Since the integrals in Eq.~(\ref{eq:cutler_mott}) are dominated by a single resonance, the reduction of the denominator for non-vanishing $\sigma$ is compensated by the analog reduction of the nominator. Only for $\sigma / \delta \geq 0.2$, the Seebeck peak starts to diminish and another peak arises at higher temperature which originates from the next conductance resonance.

In Fig.~\ref{fig:conduct_seebeck}(c), the chemical potential is between the first and second resonance but closer to the lower than the upper one.
As a consequence, at low temperatures $\SG$ is first dominated by the lower resonance which yields a positive thermopower. With growing $ T $ the influence of the broader upper resonance appears and eventually turns $\SG$ even negative. Interestingly, in this higher temperature regime even a rather pronounced peaks can appear for growing $ \sigma $. The reason lies in the different suppression of the resonance peaks in $ \langle G \rangle $. The upper resonance is more slowly reduced, see Fig.~\ref{fig:conduct_seebeck}(a). 

\subsection{Analytical aspects of the thermopower $\SG$}

\subsubsection{Validity of Mott's formula}

The standard approximation for the thermopower Eq.~(\ref{eq:cutler_mott}) in the low-temperature regime is Mott's formula
\begin{equation}
\SG  = -\frac{\pi^2}{3} \frac{k_B^2 T}{e} \left.\frac{\partial \log \langle G \rangle }{\partial E}\right|_\mu \; ,
\label{eq:mott-2}
\end{equation}
which is valid at low temperatures where the thermopower is dominated by low energy excitations of the charge carriers at the chemical potential $\mu$.
We use now Eq.~(\ref{eq:g-approx}) and obtain with Eq.~(\ref{eq:mott-2}),
\begin{equation}
\SG  = -\frac{\pi^2}{3} \frac{k_B^2 T}{e} 2N \left. \frac{\partial \log |t_1|  }{\partial E}\right|_\mu = - S_0 \frac{2\pi^2}{3}  \frac{N k_B T}{E_c}   \; ,
\label{eq:s-tn}
\end{equation}
where we defined a characteristic energy scale $ E_c $ by
\begin{equation}
\left. \frac{\partial \log |t_1|  }{\partial E}\right|_\mu = \left. \frac{\partial |t_1|}{\partial E} \frac{1}{|t_1|} \right|_{\mu} = \frac{1}{E_c}  \; .
\end{equation}
Note that $ E_c $ contains the information about the conductance $ \langle G \rangle $ as well as the position of the chemical potential. 
In Eq.~(\ref{eq:s-tn}) we observe the surprising result that in the Mott regime the Seebeck coefficient is proportional to $ N $, a result which had been previously obtained numerically \cite{guttman_thermoelectric_1995}. Thus, the higher the number of impurities the better the thermoelectric performance is found. 

\begin{figure}[ht]
\center
\includegraphics[scale=1.0]{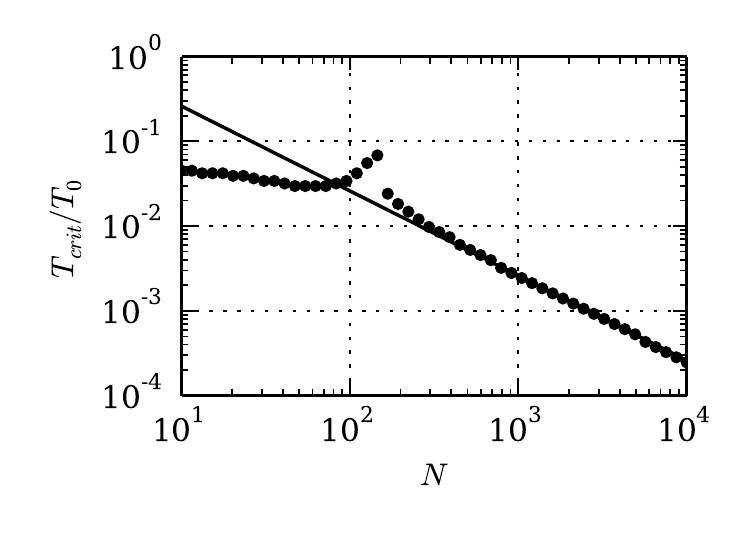} 
\caption{Numerical calculation of the critical temperature as a function of the number of impurities~$N$ (dots) and a $1/N$ fitting function (solid line) for $\tilde{\mu} = \mu/V_0 = 4,N=1000,\lambda=9$.}
\label{fig:T_crit}
\end{figure}

The question arises up to which temperature the Mott regime is a valid approximation. Therefore we define a critical temperature $ T_{crit} $ which we consider as an upper bound for the Mott regime. We calculate $ \SG $ with Eq.~(\ref{eq:cutler_mott}) as well as Mott's formula and determine $T_{crit}$, the temperature at which the relative discrepancy exceeds 5\%, as a function of $N$. The result is displayed in Fig.~\ref{fig:T_crit} as a double-logarithmic plot, showing that $ T_{crit} \propto N^{-1} $. 

We can argue on this $N$-dependence of $ T_{crit} $ by the following discussion. Since in the integrals only the energy range close to $ \mu $ is important at low-temperature, we may expand $ t_1(E) $ around $ \mu $ assuming a weak energy dependence,
\begin{equation}
t_1(E) \approx t_1(\mu) \left[1 + \left( \frac{E - \mu}{E_c} \right) \right] \; ,
\end{equation}
which yields
\begin{equation}
\langle G \rangle \approx 2 G_0 e^{2N \ln |t_1(\mu)| + 2N (E-\mu)/E_c} \;.
\end{equation}
This inserted into Eq.~(\ref{eq:cutler_mott}) gives
\begin{equation}
\SG = - \frac{S_0}{k_B T} \frac{K_1}{K_0} \; ,
\end{equation}
with
\begin{eqnarray} 
K_n & = & \int dE (E-\mu)^n \langle G \rangle(E) \left(- \frac{\partial f}{\partial E} \right) \nonumber\\ 
& \approx  & \frac{(k_B T)^n}{4} A \int_{-\infty}^{+\infty} dx \; x^n \frac{e^{2x \gamma}}{\cosh^2(x/2)} \; ,
\end{eqnarray}
where $ \gamma = N k_B T / E_c $ and $ A = 2 G_0 e^{2N \ln | t_1(\mu)|} $. Obviously, the low-temperature limit is only well defined if
\begin{equation}
\gamma = \frac{N k_B T}{E_c} < \frac{1} {2} \; ,
\end{equation}
such that the upper limit can be estimated as
\begin{equation} 
T_{crit} \sim \frac{E_c}{k_B N} \; ,
\end{equation}
as obtained numerically. It also indicates that the validity of Mott's formula is restricted to the very low temperature regime which shrinks with the system size,
if localization plays a role. In order too discuss the high temperature regime and the influence of the conductance resonances we will have to go beyond Mott's approach.

\subsubsection{Beyond Mott's formula: the $\delta E$-expansion}
\label{sec:beyond_Mott}

Mott's formula is inappropriate to treat the extrema of thermopower observed in our calculation displayed in Fig.~\ref{fig:conduct_seebeck}(b,c). We resort therefore to a different approach.  For simplicity, we focus on a single conductance peak at the resonance energy $E_n$ for $\sigma = 0$ and expand the expression in Eq.~(\ref{eq:aver_trans_perfect}) near $E_n$, 
\begin{equation}
 \langle G \rangle \approx \frac{2 G_0}{1 + \left( 1+ \Gamma_n (E-E_n)^2  \right)^N} \; ,
\end{equation}
which has a width at half maximum of ${\delta E \approx \sqrt{8/\Gamma_n N}}$. Note that $ \delta E \propto N^{-1/2} $ indicates that localization yields sharper resonances as states beyond the resonance energy tend to be localized. 

\begin{figure}[ht]
\center
\includegraphics[scale=1.0]{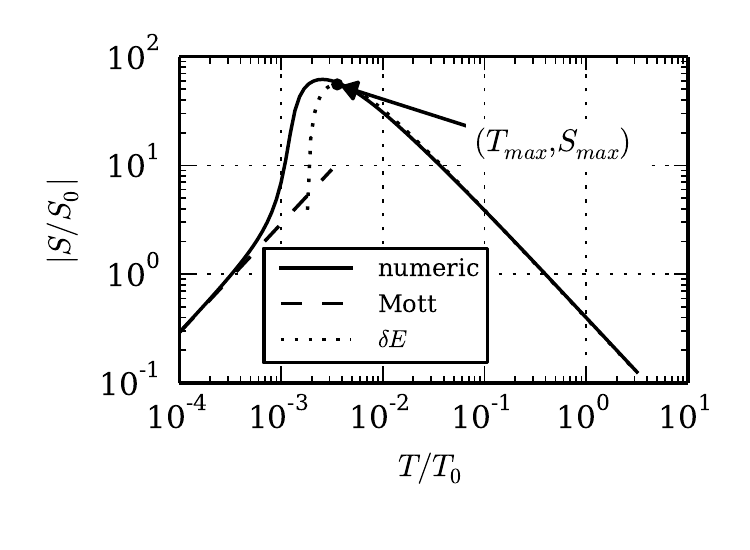} 
\caption{The low temperature expansion (Mott) and the $\delta E$~expansion in direct comparison with numerical data of $S_{\langle G \rangle}$ for $\lambda = 9, N=1000, \mu/V_0=1.7$.}
\label{fig:seebeck_bothapprox}
\end{figure}

Since $\delta E \rightarrow 0$ as $N \rightarrow \infty$ we will use $\delta E$ as an expansion parameter and express the Seebeck coefficient as
\begin{eqnarray}
 \SG  & \approx&  - \frac{1}{e T} \left[ \Delta E + \frac{\Lambda}{8} \delta E^2 \frac{\partial}{\partial E} \log \left( -\frac{\partial f}{\partial E} \right)_{E_n}  \right] \nonumber\\
 & = & -S_0 \left[ \frac{\Delta E}{k_BT} - \frac{\Lambda}{8} \frac{\delta E^2}{(k_B T)^2} \tanh \left(\frac{\Delta E}{2 k_BT} \right) \right]  ,
 \end{eqnarray}
where $\Delta E = E_n - \mu$ is the distance between the resonance energy and the chemical potential, and $\Lambda \approx 0.6324$ (see Appendix \ref{app:delta_E:general},\ref{app:delta_E:thermopower} for a detailed derivation). From this expression it becomes obvious that the sign of $ \SG $ depends on the sign of $ \Delta E $, i.e. whether the chemical potentials is above or below the resonance. As Fig.~\ref{fig:seebeck_bothapprox} shows, the $\delta E$-expansion matches perfectly the high temperature limit of $\SG(T)$ and is also able to include qualitatively the thermopower extremum which is invisible within Mott's formula. 
The extremum can easily be discussed by using the following approximation, 
\begin{equation}
 \SG(T) \approx - S_0 \left( \frac{|\Delta E|}{k_B T} - \frac{\Lambda}{8} \frac{\delta E^2}{(k_B T)^2} \right) \cdot \textrm{sign}(\Delta E).
\end{equation}
The computation of the extremal point is then straightforward,
\begin{eqnarray}
k_B T_{max} &=& \frac{\Lambda}{4} \frac{\delta E^2}{|\Delta E|}, \label{eq:T_max} \\
\frac{S_{max}}{S_0} &=& -\frac{2}{\Lambda} \left( \frac{\Delta E}{\delta E} \right)^2 \cdot \textrm{sign}(\Delta E).
\label{eq:delta_E_expansion_results}
\end{eqnarray}
Note that $ S_{max} \propto N $, indicating that localization tends to increase the thermoelectric performance due to the sharpening of the resonance peaks. 
It is interesting to remark that within our approach the product
\begin{equation}
 |T_{max} \cdot S_{max}| = \frac{|\Delta E|}{2 e}
\end{equation}
becomes independent of $N$, the number of impurities. 

\begin{figure*}[ht]
\center
\includegraphics[scale=1.0]{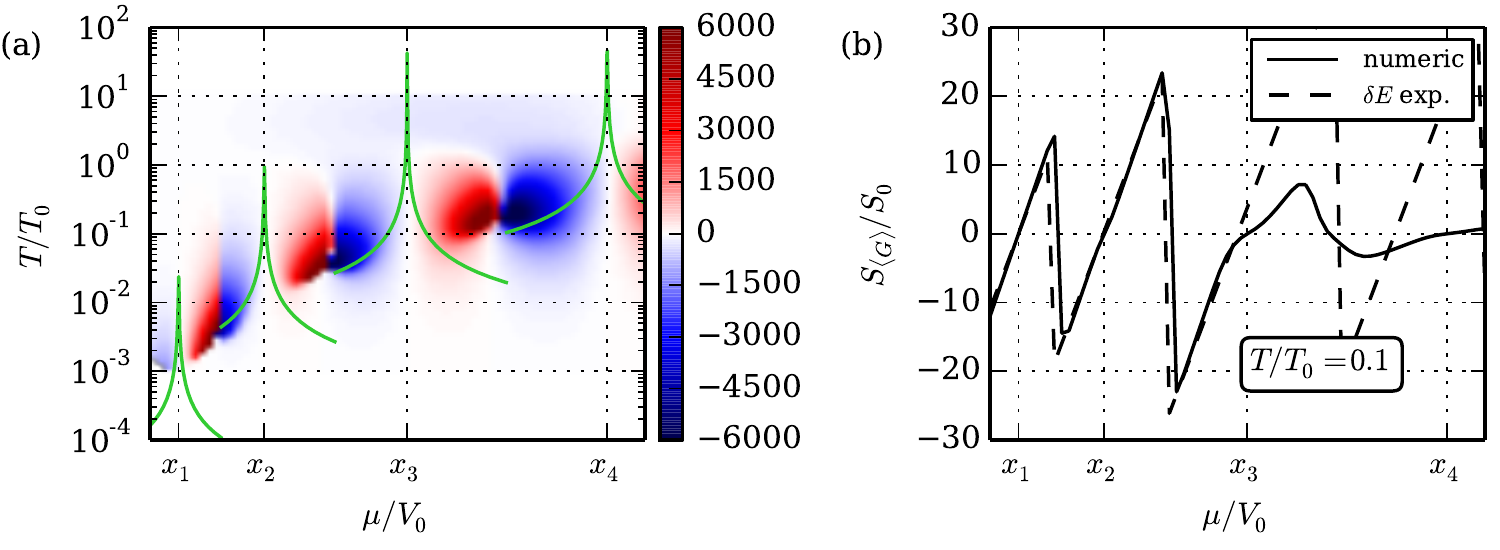}
\caption{(Color online) (a) We plotted the thermopower $S$ multiplied by $\mu^{2.4}$ such that the extrema are equally pronounced for $\lambda=9,N=8000$. In (b) we plotted the thermopower for the temperature $T/T_0 = 0.1$ which demonstrates that the $\delta E$ expansion is in good agreement with the numerical data as long as $T > T_{crit}$ which is true for the lowest two conductance peaks.}
\label{fig:seebeck_regionplot}
\end{figure*}

\subsubsection{Several conductance peaks}

The statements based on the $\delta E$-expansion can be extended to the multi-peak scenario. Let us focus on low temperatures, $k_B T < E_{n+1}-E_n$, where the thermopower shows a single extremum. It corresponds to the conductance peak which dominates the integrals in Eq.~(\ref{eq:cutler_mott}) and lies closest to the chemical potential $\mu$.

In Fig.~\ref{fig:seebeck_regionplot}(a) we plot the numerical evaluation of the Seebeck coefficient $S$ as a function of $T$ and $\mu$. For illustration purpose we multiply $S$ with the factor $\mu^{2.4}$ such that the extrema are equally pronounced. The green line describes the temperature $T_{max}$ of the extrema as a function of $\mu$, according to Eq.~(\ref{eq:T_max}). The locations of diverging $T_{max}$ mark the resonance energies. In addition, we display a cut along $T/T_0 = 0.1$ to show $S$ as a function of $\mu$ in Fig.~\ref{fig:seebeck_regionplot}(b). We can here compare the numerical and the approximate result of Eq.~(\ref{eq:delta_E_expansion_results}) using always the closest conductance peak.
The numerical data support the validity of a single peak picture, if the chemical potential is in the vicinity of a conductance peak. Then the thermopower shows a linear behavior with a sign change as a function of $\mu$, in agreement with the analytical results for $T>T_{max}$ (for $T<T_{max}$ the $\delta E$-expansions breaks down). Roughly in the middle between two resonance energies, there is an abrupt change of sign of the thermopower which corresponds to the switch in the dominance of the adjacent resonances. 

The temperatures $T_{max}$ of the individual conductance peaks are qualitatively in agreement with the numerical data, although there are discrepancies due to the influence of neighboring resonances, which is more pronounced for chemical potentials above the dominant resonance.  

\subsection{$\delta E$-expansion of the $ZT$-value}

The figure of merit $ZT$ of a thermoelectric device describes its efficiency. It is given by
\begin{equation}
 ZT = T \frac{\sigma_{el} S^2}{\kappa_{el} + \kappa_{ph}}
\end{equation}
with the electrical conductivity~$\sigma_{el}$, the electronic and phonon heat conductivity~$\kappa_{e}$ and $\kappa_{ph}$, respectively. 

If we simply neglect $\kappa_{ph}$, the $\delta E$-expansion of $ZT$ can be straightforwardly evaluated (see Appendix \ref{app:delta_E:ZT}) which in leading order in $\delta E$ gives, 
\begin{equation}
ZT \approx \frac{8}{\Lambda} \left( \frac{\Delta E}{\delta E} \right)^2 \propto N .
\label{eq:zt_deltaE}
\end{equation}

\begin{figure*}[htb]
\center
\includegraphics[scale=1.0]{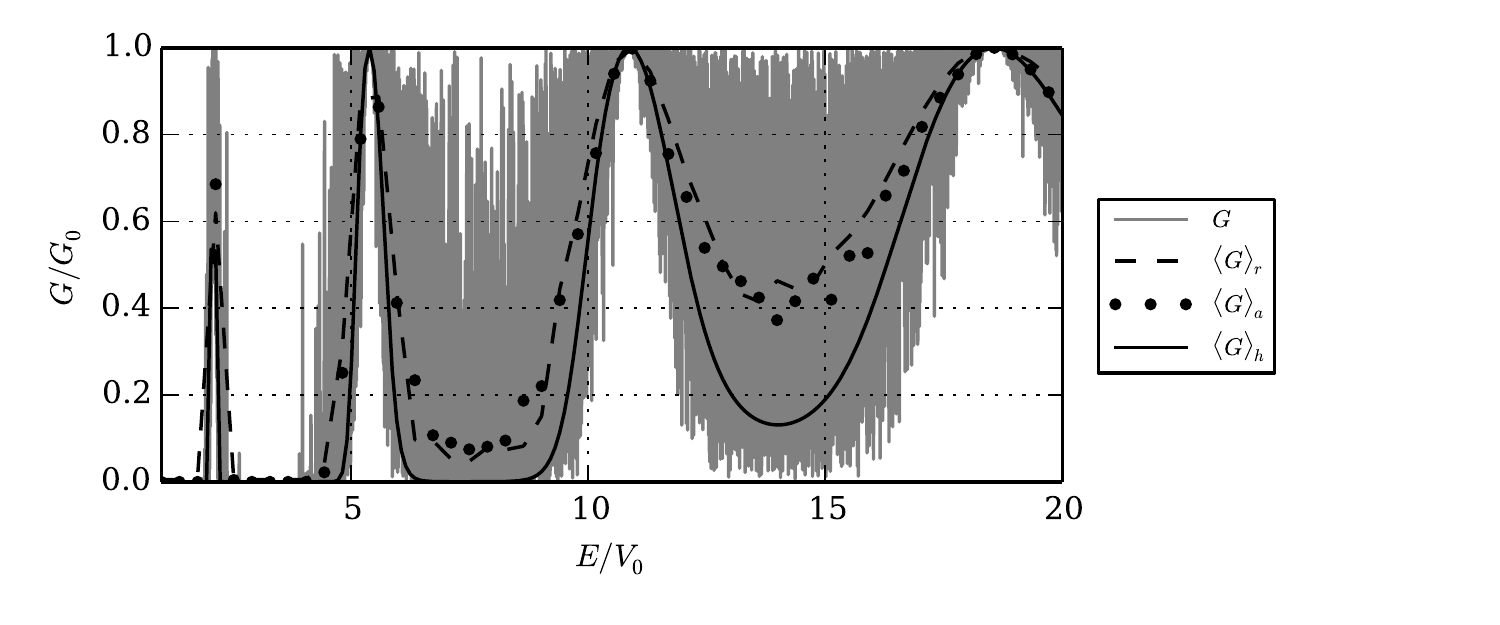}
\caption{A comparison of the arithmetic, harmonic, renormalized averaged conductance and the numerically calculated conductance.}
\label{fig:num_average}
\end{figure*}

This result is in agreement with numerical calculations of $ZT$ as long as $T > T_{max}$. The apparently unlimited range of $ZT$ values is a consequence of $\kappa_{ph}=0$, because the dependence on $\delta E^{-2}$ originates from ${\kappa_{el} \propto \delta E^2}$.
By taking the phonon heat conductance into account the $ZT$-value stays limited for all $\delta E$ and $\Delta E$.  We parameterize the phonon heat conductance as
\begin{equation}
\kappa_{ph}(T) = \left( \frac{k_B}{e} \right)^2 \frac{\Lambda_2 G_0}{8 \sqrt{8}} T_{ph}(T) = \alpha T_{ph}(T) \; ,
\end{equation}
where $\alpha = 3.356 \cdot 10^{-14}\; W/K$ and $ T_{ph}(T) $ describes the temperature dependence.  A straightforward calculation leads to the maximum of the $ZT$-value with respect to $\delta E$ and $\Delta E$ for given temperature $ T$,
\begin{equation}
(ZT)_{max} = \underset{\delta E,\Delta E}{\textrm{max}} ZT \approx \gamma \left( \frac{T}{T_{ph}(T)} \right)^{2/3} \; ,
\end{equation}
where $\gamma = 7.696$. The optimal value for $\delta E$ and $\Delta E$ which maximizes the $ZT$-value are given by
\begin{eqnarray}
 \Delta E &\approx& \pm 3.24 \cdot k_B T \; , \\
 \delta E &\approx& 0.417 \cdot k_B T \left( \frac{T_{ph}(T)}{T} \right)^{1/3} \; .
\end{eqnarray}
This optimal value for $\Delta E$ is qualitatively in agreement with the results by Mahan and Sofo \cite{mahan_best_1996} who performed similar computations for systems described by the Boltzmann transport theory. They found an optimal value of $\Delta E = 2.4 k_B T$ for a $\delta$-function shaped transport distribution function (mobile density of states), i.e. for $\delta E = 0$. This discrepancy arises from the upper boundary $G_0$ of the conductance $G$ which prevents a $\delta$-function shaped conductance $G$ and yields a finite optimal width of the peak, $\delta E \neq 0$, and a larger coefficient for $\Delta E$.

\begin{figure}[htb]
\center
\includegraphics[scale=1.0]{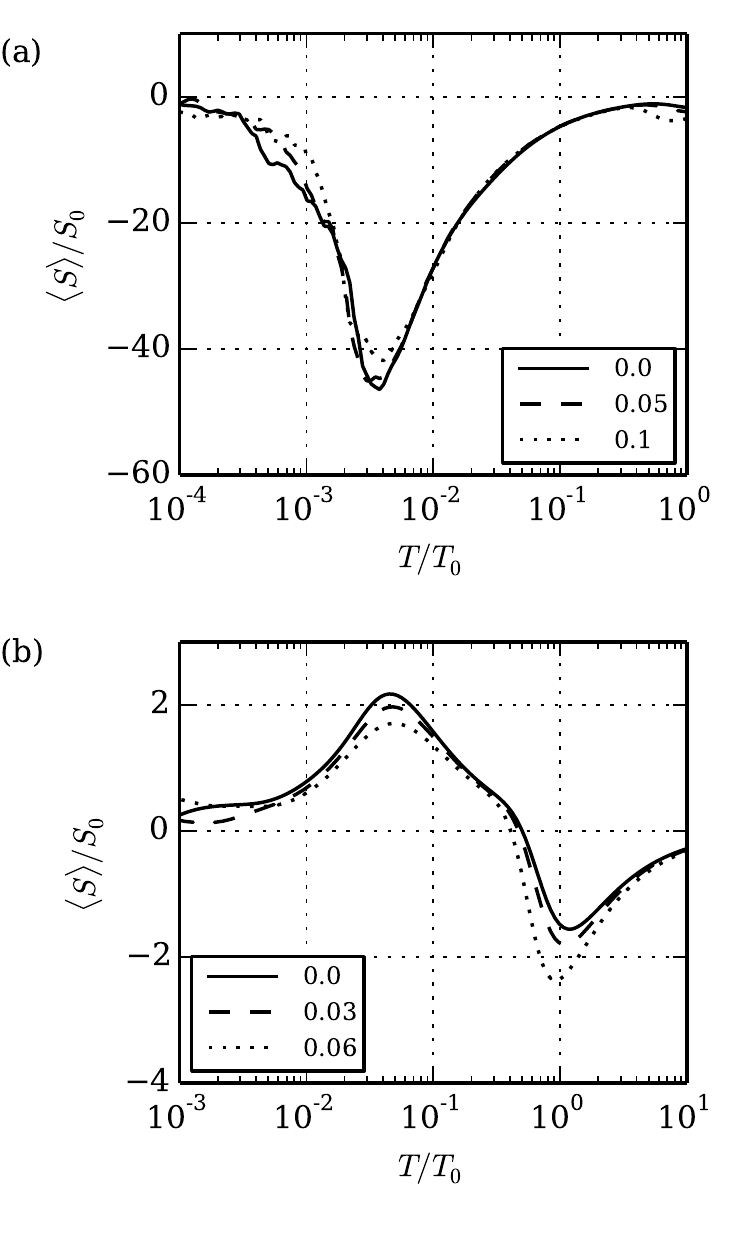}
\caption{Numeric results of $\langle S \rangle$ for (a) $\tilde{\mu} = \mu/V_0 = 1.6$ (slightly below the first resonance) and (b) $\tilde{\mu} = 2.3$ (between the first and the second resonance) with $N=1000$ and $\lambda=9$, see Fig.~\ref{fig:conduct_seebeck}.}
\label{fig:num_sim}
\end{figure}

\section{Numerical Simulations}
\label{sec:Numerical_Simulation}

The Seebeck coefficient $\SG$ derived via the harmonically averaged conductance $\langle G \rangle_h$ can be treated analytically in various limits, in particular, also using the $\delta E$-expansion. However, the harmonic average of the conductance $\langle G \rangle_h$ takes only the broad resonance peaks due to resonant scattering in the impurity barrier potential into account. It underestimates the contributions from the narrow peaks due to resonant scattering between the impurities, as we illustrate in Fig.~\ref{fig:num_average}. We compare the different averages with the numerical calculation of the conductance for a concrete configuration which involves numerous narrow resonances. The arithmetic average does much better in this respect.

\subsection{Validity of the averages for $S$}

In order to understand this discrepancy qualitatively we define the renormalized averaged conductance $\langle G \rangle_{r}$. This averaged conductance is based on the harmonic average $\langle G \rangle_{h}$, but includes the influence of the very narrow peaks between the broad resonances. 
We introduce the mean distance $\Delta$ between neighboring narrow peaks and their mean width $\omega$. Since the harmonic average is dominated by the smallest values, $\langle G \rangle_h$ can be regarded as the minimal conductance, consistently with Fig.~\ref{fig:num_average}. To this minimal value we should add the contributions of the narrow resonances, each one with the mean weight $w=C \cdot \omega \cdot (G_0-\langle G \rangle_h)$ where $ C \sim 1 $ is a numerical factor. The renormalized harmonic average $\langle G \rangle_{r}$ takes all those small weights into account by distributing them uniformly, 
\begin{equation}
 \langle G \rangle_{r} = \langle G \rangle_h + C \frac{\omega}{\Delta} (G_0-\langle G \rangle_h).
 \label{eq:renormalized_conductance}
\end{equation}
Indeed Fig.~\ref{fig:num_average} shows that the renormalized average $ \langle G \rangle_{r}$  is much closer to the arithmetical average $ \langle G \rangle_{a}$ than the harmonic average. Thus we would expect that this also leads to numerically more accurate results for the Seebeck coefficient.

For this discussion let us consider the numerically determined Seebeck coefficient $ \langle S \rangle $ and compare it with our previous averaging procedure based on the harmonic average of the conductance. We analyze the same regimes as in Sect.\ref{sect:s-aver} at $\lambda=9,N=1000$. We start with 
$\mu/V_0 = 1.6 $ which yields a single thermopower extremum in the chosen temperature range. For the numerical average we used 50 samples randomly picked according to the distribution function in Eq.~(\ref{eq:distribution}), calculated the conductance and the Seebeck coefficient by means of the Cutler-Mott formula Eq.~(\ref{eq:cutler_mott}) which then is averaged over all samples. We can compare now the results for $\langle S \rangle$ in Fig.~\ref{fig:num_sim}(a) with the analogous results for $S_{\langle G \rangle}$ in Fig.~\ref{fig:conduct_seebeck}(b) obtained with $\langle G \rangle_h$. Obviously, the qualitative features are identical: there is a maximum at roughly the same temperature and also the trends as a function of the standard deviation $ \sigma $ agree. The largest difference lies in the absolute magnitude, $ \SG $ is roughly three times the size of $ \langle S \rangle $. 

The discrepancy can be accounted for by analyzing the consequences of the renormalized average $\langle G \rangle_r$. A straightforward calculation leads to a renormalized width of the conductance peak,
\begin{equation}
\delta E_{ren} = \frac{\delta E}{\sqrt{1-2 C \frac{\omega}{\Delta}}} \; ,
\end{equation}
which can used to determine the influence on the extremum by using Eq.~(\ref{eq:T_max},\ref{eq:delta_E_expansion_results}),
\begin{equation}
T_{max}^{ren}=\frac{T_{max}}{1-2C \frac{\omega}{\Delta}} \; , \; S_{max}^{ren}=S_{max} \left( 1-2C \frac{\omega}{\Delta} \right).
\end{equation}
From the numerical calculation we estimate $ C \omega / \Delta \sim 0.3 - 0.37 $ in the vicinity of the first resonance, which can account for the observed reduction of $ \langle S \rangle $ relative to $ \SG $. Moreover, the shift of $ T_{max} $ is consistent as well. 

We consider now the second case where several (broad) conduction resonances of the barriers are involved, by choosing the chemical potential $\mu/V_0=2.3$ and the same averaging procedures. The result for $\langle S \rangle$ can be seen in Fig.~\ref{fig:num_sim}(b) which can be compared with the results from $S_{\langle G \rangle}$ in Fig.~\ref{fig:conduct_seebeck}(c). Again the characteristic behavior is the same in both plots. We see a sign change in the same temperature range and also the overall modification upon increasing $ \sigma $ have the same direction. Again the overall magnitude is reduced by a factor 3 in $ \langle S \rangle $ with respect to $ \SG $.

\subsection{Fluctuations of $ \langle S \rangle $}

\begin{figure}[h]
\center
\includegraphics[scale=1.0]{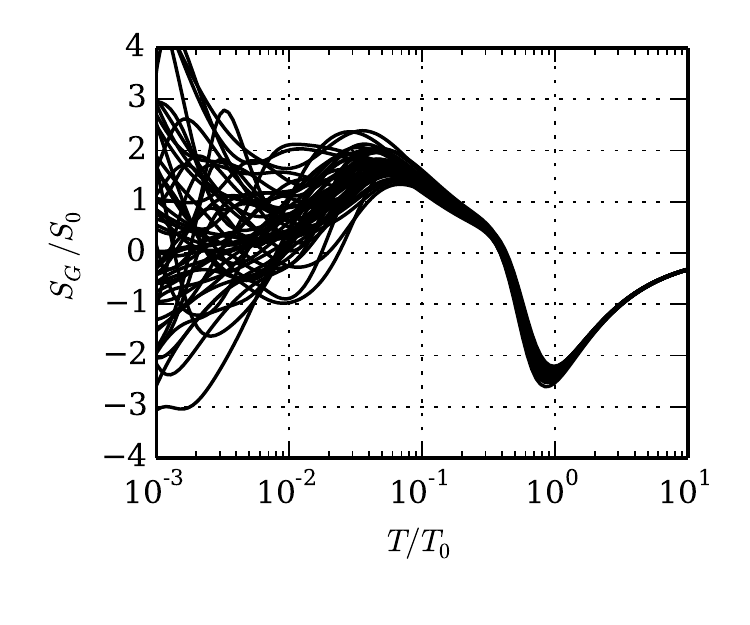} 
\caption{A collection of Seebeck coefficients $S_{G_i}$ as a function of temperature which correspond to 50 different realizations of the underlying parameters for $N=1000,\mu/V_0 = 2.3,\lambda=9,\sigma/\delta = 0.06$.}
\label{fig:Seebeck_ensemble}
\end{figure}

Within our approach to $\langle S \rangle$ it is also of interest to examine the fluctuations at low temperature. In Fig.~\ref{fig:Seebeck_ensemble} we plot the thermopowers corresponding to 50 configuration samples for $N=1000,\mu/V_0=2.3,\lambda=9$. At temperatures $T/T_0 > 10^{-1}$, the thermopower is confined in a very narrow range where the common broad (impurity) resonances determine the thermopower's behavior. As we decrease temperature, the thermopowers starts to reflect more and more the individual nature of different configurations, i.e. the influence of the tiny (inter-impurity) resonances is increasing. For $T/T_0 \lesssim 10^{-3}$ the different thermopowers are spread over a broad range and the averaged thermopower is almost zero, see Fig.~\ref{fig:num_sim}(b), i.e. the tiny inter-impurity resonances dominate the thermopower's behavior.

\begin{figure}[h]
\center
\includegraphics[scale=1.0]{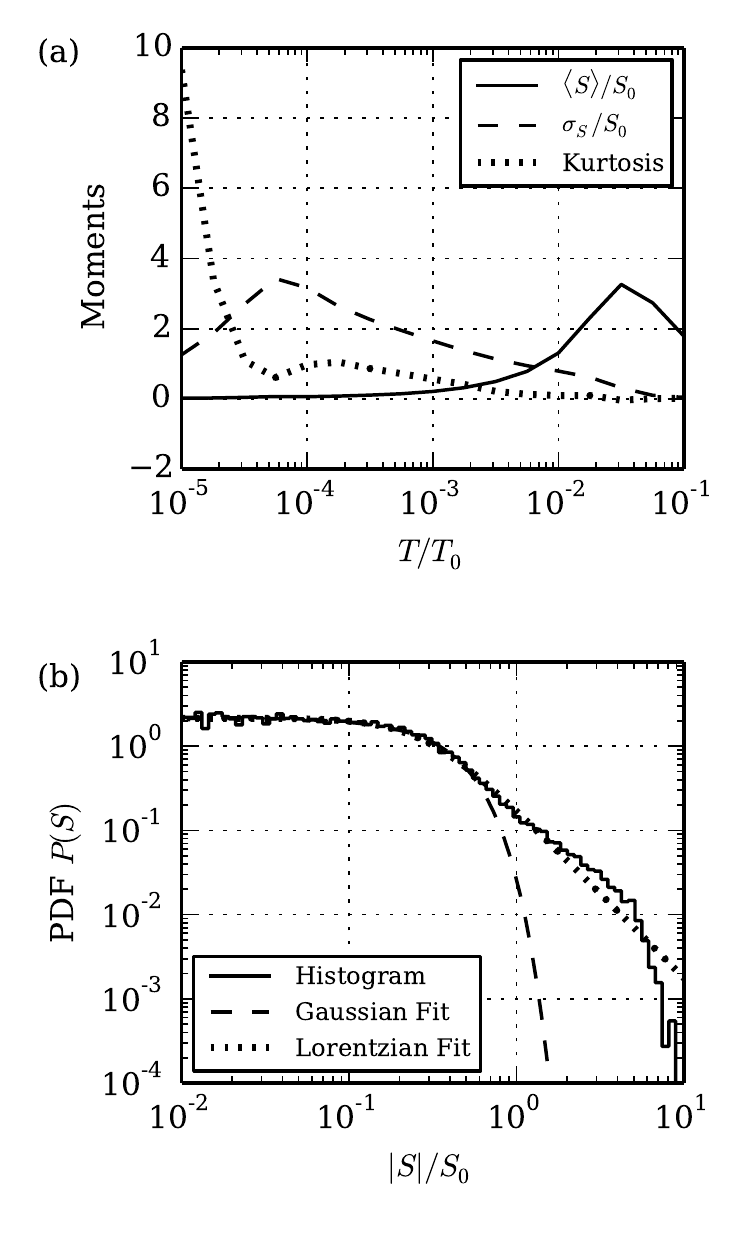} 
\caption{(a) Plot of the average $\langle S \rangle / S_0$, the standard deviation $\sigma_S / S_0$ and the excess kurtosis. (b) Comparison of the probability distribution function $P(S)$ for $T/T_0 = 10^{-5}$ with a Gaussian and a Lorentzian fit. The statistics involves 50'000 samples for $N=2000,\mu/V_0=2.3,\lambda=9, \sigma/\delta = 0$. }
\label{fig:NumData_overview}
\end{figure}

In order to characterize the fluctuations in terms of the shape of the probability distribution function we computed 50'000 samples of $S$ for $10^{-5} < T/T_0 < 10^{-1}$ with $N=2000,\mu/V_0=2.3,\lambda=9,\sigma/\delta=0$ and determined the mean $\langle S \rangle / S_0$, the standard deviation $\sigma_S/S_0$, the skewness $\langle ( S - \langle S \rangle)^3  \rangle / \sigma_S^3$ and the excess kurtosis $\langle (S-\langle S \rangle)^4 \rangle/\sigma_S^4-3$ (vanishing for a Gaussian distribution). The results are shown in Fig.~\ref{fig:NumData_overview}(a) where we skipped the skewness since it fluctuates in a very narrow regime ($<0.1$) around zero.

In the temperature range $10^{-2.5} < T/T_0 < 10^{-1}$, the small skewness and the vanishing excess kurtosis suggest a Gaussian shaped probability distribution function for $S$. When $T/T_0$ approaches $10^{-2.5}$ the standard deviation increases and the finite average $\langle S \rangle$ decreases which demonstrates that the tiny inter-impurity resonances have a growing influence on the thermopower (i.e. $k_B T \gtrsim \Delta$) whereas the broad impurity resonances are loosing their impact.

In the regime where $10^{-4.25} < T/T_0 < 10^{-2.5}$, the vanishing average $\langle S \rangle$ and the finite standard deviation reveals that the behavior of the thermopower is entirely dictated by inter-impurity resonances ($k_B T < \Delta$). The finite excess kurtosis shows that the tails of the probability distribution function becomes longer compared to the Gaussian distribution function. However, the detailed nature of the distribution could not be clarified so far. 

At very low temperatures, ${10^{-5} < T/T_0 < 10^{-4.25}}$, the excess kurtosis is strongly enhanced, i.e. the tail must have considerably been stretched. In combination with the reduction of the standard deviation, we conclude that the nature of the probability distribution function has profoundly changed from a rather Gaussian shape to a widely spread shape. In Fig.~\ref{fig:NumData_overview}(b) we plotted the numerical probability distribution function including a Lorentzian and a Gaussian fit on a log-log scale. It is evident that the probability distribution is in agreement with the Lorentzian distribution function for ${10^{-0.5} < |S|/S_0 < 10^{0.75}}$ which is consistent with the results by van Langen et al \cite{van_langen_thermopower_1998}.

\section{Summary}
\label{sec:discussion}

With our study we target some of the basic properties of a specially designed disordered finite-size system which shows two main 
features, Anderson localization due to random scattering and conductance peaks introduced by resonant transmission because of similarly shaped barriers.
We show that localization plays an important role, as it sharpens the conductance resonances. This in turn manifests itself in the enhancement of 
thermopower. One feature is the $ S \propto N T $ behavior at very low temperatures which indicates an enhancement in finite systems with stronger 
localization\cite{guttman_thermoelectric_1995}. We also found that the maximum of $|S|$ in an intermediate temperature range is again proportional to $ N $, if a single resonance dominates the behavior. Eventually, the figure of merit could be boosted as well by localization effects ($ ZT \propto N $), and is only truncated by the phonon contribution to the heat transport. 

Within the Landauer-B\"uttiker formalism we developed a procedure based on the harmonic average of the conductance $G$ to determine the mean value of the Seebeck coefficient, $ \SG $. Examining the validity of this scheme in comparison with numerical results, we could show that basic results can be trusted, while we could identify the short-comings. We also could determine the range of validity for Mott's formula and found a good approximation for the intermediate temperature range introducing a scheme in the expansion in $ \delta E $, the width of the resonance. 

Although our model has rather severe constraints we believe that techniques in producing one-dimensional samples with special design might soon reach a stage where it could be emulated. We have also demonstrated that less perfect scatterers might still give a good performance.


\section{Acknowledgement}

We are grateful to B. Batlogg, J. Buhmann, D. Ivanov, M. Ossadnik, S. Poppulo and A. Weidenkaff for helpful discussions. This study was supported by the Swiss Nationalfonds, the NCCR MaNEP, the HITTEC project of the Competence Center Energy \& Mobility and the Sinergia TEO, as well as by the Toyota Central R \& D Laboratories, Nagakute, Japan.

\appendix

\section{Computation of the harmonic average}
\label{app:averaging_process}

\subsection{General formula for the harmonic average $\langle G \rangle_h$}
\label{app:averaging_process:result}

In this Appendix we address to the problem of calculating the harmonic average of the conductance $\langle G \rangle_h$, see Eq.~(\ref{eq:t_average}), for the situation where the transfer matrices of the \textit{elastic} events (scattering and propagation) are known and the probability distribution function $P$ decomposes into single event probability distribution functions $P_{i,p}$, i.e.
\begin{equation}
 P = \prod_j  P_p(l_j) P_i(\delta_j,V_j) \; .
\end{equation}
The naive calculation by taking the square of the absolute value of the entity of $\hat{T}$ and average it afterwards normally ends in a recursive formula which might be very challenging to solve.
Thus, we would like to access the problem in a different way and note that
\begin{equation}
	\frac{1}{2} \textrm{Tr} \left( \hat{T} \hat{T}^\dagger \right) = \frac{1+|r|^2}{|t|^2} = \frac{2}{|t|^2} - 1.
\end{equation}
We introduce the harmonic average
\begin{equation}
 \langle \hat{T}\hat{T}^\dagger \rangle = \int \left( \prod_j dl_j d\delta_j dV_j \ P_p(l_j)P_i(\delta_j,V_j)\right) \hat{T} \hat{T}^\dagger
 \label{eq:ttdagger}
\end{equation}
and find by comparing with Eq.~(\ref{eq:t_average}) that
\begin{equation}
 \langle G \rangle_h = \frac{2 G_0}{1+ \frac{1}{2}\textrm{Tr} \langle \hat{T} \hat{T}^\dagger \rangle}.
\end{equation}
The transfer matrix $\hat{T}$ of a system with $N$ impurities can be decomposed such that we can write
\begin{eqnarray}
 &&\hat{T}\hat{T}^\dagger\Big|_{N} = \nonumber \\  && \hat{T}_p(l_N) \hat{T}_i(\delta_N,V_N)  \left[ \hat{T}\hat{T}^\dagger\Big|_{N-1} \right]
  \hat{T}_{i}^\dagger(\delta_N,V_N) \hat{T}_p^\dagger(l_N) \; . \nonumber\\  
 \label{eq:ttdagger2}
\end{eqnarray}

We define a mapping $f$ on the vector space of 2x2-matrices
\begin{eqnarray}
 f(\hat{X}) &:=& \int dl \; dV \; d\delta \ P_p(l) P_i(\delta,V) \  \nonumber \\
 &&  \hat{T}_p(l) \hat{T}_i(\delta,V) \hat{X}\ \hat{T}_i^\dagger(\delta,V)\hat{T}_p^\dagger(l) \; ,
 \label{eq:f}
\end{eqnarray}
which allows us to write
\begin{equation}
\langle \hat{T} \hat{T}^\dagger \rangle_N = f\left( \langle \hat{T} \hat{T}^\dagger \rangle_{N-1} \right) = \dots = f^N ( \mathbbm{1} ) \; ,
\end{equation}
where we used that $\langle \hat{T} \hat{T}^\dagger \rangle_{N=0} = \mathbbm{1}$.

The function $f$ is a linear mapping and can therefore be fully described by a matrix. We choose the Pauli matrices $ \tau_2, \tau_3, \tau_4 $ combined with the identity matrix~$\tau_1$ 
\begin{equation}
 \tau_1 = \mathbbm{1}, \ 
 \tau_2 = \begin{pmatrix} 0 & 1 \\ 1 & 0 \end{pmatrix}, \
 \tau_3 = \begin{pmatrix} 0 & -i\\ i & 0 \end{pmatrix}, \
 \tau_4 = \begin{pmatrix} 1 & 0 \\ 0 & -1\end{pmatrix}
\end{equation}
as a basis of the vector space. With
\begin{equation}
 \hat{X} = \sum_{j=1}^4 X_j \tau_j \ , \quad \hat{Y} = f( \hat{X} ) = \sum_{j=1}^4 Y_j \tau_j
 \label{eq:basis_expand}
\end{equation}
the linear mapping $f$ is fully described by the 4x4 matrix~$\hat{f} = \{ f_{ij} \}$,
\begin{equation}
 Y_i = \sum_{j=1}^4 f_{ij} X_j.
\end{equation}
The matrix $\hat{f}$ is then computed by using Eq.~(\ref{eq:f}) and is given by
\begin{equation}
 f_{kl} = \sum_{j=1}^4 f_{kj}^p f_{jl}^i.
\end{equation}
The matrix $\hat{f}^{p}$ depends only on the probability distribution $P_p$ and the transfer matrix $\hat{T}_p$. By parameterizing 
\begin{equation}
 \hat{T}_p = \begin{pmatrix} T_{11} & T_{12} \\ T_{12}^* & T_{11}^* \end{pmatrix} \; ,
 \label{eq:param_T}
\end{equation}
we find that
\begin{equation}
 \hat{f}^p = \left(
 \begin{array}{ccc|c}
   &         &    & 0 \\
   &   \hat{F}_p   &    & 0 \\
   &         &    & 0 \\
   \hline
 0 &   0     & 0  & \gamma_p 
 \end{array}
 \right) \; ,
 \label{eq:matrixF}
\end{equation}
where the 3x3 matrix $\hat{F}_p$ is
\begin{eqnarray}
 && \hat{F}_p = \int \ dl \ P_p(l) \times \nonumber \\
 && \begin{pmatrix}
  |T_{11}|^2 + |T_{12}|^2   & 2 \Re(T_{11}^* T_{12})  &  2 \Im(T_{11} T_{12}^*) \\
  2 \Re(T_{11}T_{12})       & \Re(T_{11}^2 + T_{12}^2)& \Im(T_{11}^2 - T_{12}^2) \\
  -2\Im(T_{11}T_{12})       &-\Im(T_{11}^2 + T_{12}^2)& \Re(T_{11}^2 - T_{12}^2)
 \end{pmatrix}
 \nonumber \\
\end{eqnarray}
and 
\begin{equation}
 \gamma_p = \int \ dl \ P_p(l) \Big( |T_{11}|^2 - |T_{12}|^2 \Big).
\end{equation}
The matrix $\hat{f}^i$ can be found analogously by applying the same scheme on the probability distribution function $P_{i}(\delta,V)$ and the transfer matrix $\hat{T}_i$.

Since $\textrm{Tr}(\hat{X}) = 2 X_1$, see Eq.~(\ref{eq:basis_expand}), and $\mathbbm{1} = \tau_1$ we obtain
\begin{equation}
\frac{1}{2}\textrm{Tr}\langle \hat{T}\hat{T}^\dagger \rangle = \left[ \Big( \hat{f}^p \hat{f}^i \Big)^N  \right]_{11} = \left[ \Big( \hat{F}_p \hat{F}_i \Big)^N  \right]_{11}.
\end{equation}
Thus, we finally obtain the averaged conductance
\begin{equation}
 \langle G \rangle_h = \frac{2 G_0}{1+ \left[ \left(\hat{F}_p \hat{F}_i\right)^N \right]_{11} }.
\end{equation}

\subsection{Computation of $\langle G \rangle_h$ for our model}
\label{app:harmonic_average:result}

We consider a system with the following probability distribution functions
\begin{eqnarray}
 P_l(l) &=& \textrm{const} \\
 P_V(V) &=& \delta(V-V_0) \\
 P_\delta(\delta_i) &=& \frac{1}{\sqrt{2\pi}\sigma} \cdot \exp\left[- \frac{(\delta_i-\delta)^2}{2 \sigma^2} \right].
\end{eqnarray}
With the transfer matrices Eq.~(\ref{eq:transfermatrix_propagation}) and (\ref{eq:transfermatrix_impurity}) we find the matrices
\begin{equation}
 \hat{F}_p = 
 \begin{pmatrix}
  1 & 0 & 0 \\
  0 & 0 & 0 \\
  0 & 0 & 0 
 \end{pmatrix}
  , \quad
 \hat{F}_i =
 \begin{pmatrix}
 \alpha_1 + \alpha_2 & \dots & \dots \\
 \dots & \dots & \dots \\
 \dots & \dots & \dots
 \end{pmatrix} \; ,
\end{equation}
where
\begin{eqnarray}
 \alpha_1 &=& \int \ d\delta_i \ P_\delta(\delta_i) \Big[ \cos^2(q\delta_i) + \frac{\epsilon^2}{4}\sin^2(q\delta_i) \Big] \\
 \alpha_2 &=& \int \ d\delta_i \ P_\delta(\delta_i) \Big[ \frac{\eta^2}{4} \sin^2(q\delta_i) \Big].
\end{eqnarray}
We introduce the dimensionless quantities
\begin{equation}
 x = \frac{E}{V_0} \quad , \quad \lambda = \frac{2mV_0 \delta^2}{\hbar^2}
 \label{eq:aver_trans_1}
\end{equation}
and find after some calculations
\begin{eqnarray}
 \alpha_1 + \alpha_2 &=& 1 + \frac{\sin^2 \sqrt{\lambda(x-1)}}{2x(x-1)} \nonumber \\
 &+& \left( 1-e^{-2\lambda(x-1)\sigma^2 /\delta^2} \right) \frac{\cos\left( 2\sqrt{\lambda(x-1)} \right)}{4x(x-1)}. \nonumber \\
 \label{eq:aver_trans_2}
\end{eqnarray}
Since we have
\begin{equation}
\left[ \left( \hat{F}_p \hat{F}_i \right)^N \right]_{11} = \left( \alpha_1 + \alpha_2 \right)^N \; ,
\end{equation}
due to the special shape of $\hat{F}_p$, the averaged conductance is given by
\begin{equation}
 \langle G \rangle_h = \frac{2 G_0}{1+\left( \alpha_1 + \alpha_2 \right)^N}.
 \label{eq:aver_trans_3}
\end{equation}

\section{The $\delta E$-expansion}
\label{app:1/N-expansion}

\subsection{General formula}
\label{app:delta_E:general}

In the limit $N \rightarrow \infty$, the width of the conductance peaks of $\langle G \rangle_h$ goes to zero as $1/\sqrt{N}$. We want to derive an expansion in the width of the peak, analogously to the Sommerfeld expansion, of the integral
\begin{equation}
 \int dx \ \langle G \rangle_h (x) h(x) \; ,
 \label{eq:app:integral}
\end{equation}
where we use the dimensionless variables ${x = E/V_0}$, ${\lambda = 2mV_0d^2/\hbar^2}$ and $h$ is an arbitrary function.

We concentrate on a single conductance peak at the resonance energy ${x_n = E_n/V_0 = 1+ (n \pi)^2/\lambda}$ at perfect resonance ($\sigma = 0$) and expand the averaged conductance $\langle G \rangle_h$,
\begin{equation}
 \langle G \rangle_h \approx \frac{2 G_0}{1+ \left( 1+ \gamma_n (x-x_n)^2 \right)^N} \equiv G_0 \cdot \delta_N(x-x_n) \; ,
\end{equation}
where
\begin{equation}
 \gamma_n = \frac{1}{4} \frac{\lambda^3}{(n \pi)^4 \left( 1+ \frac{(n \pi)^2}{\lambda} \right)}.
\end{equation}
For sufficiently large $N$, the main contribution to the integral Eq.~(\ref{eq:app:integral}) comes from a narrow energy interval around $x_n$, which we assume not to be close to $1$. 
Then we perform a Taylor expansion of $h$ about $x_n$ and extend the lower integral boundary to infinity. The integral is then
\begin{eqnarray}
 &&\int_{-\infty}^{\infty}   dx \ \delta_N(x-x_n) h(x) \nonumber \\
  && = A_0 \; h(x_n) + A_1 \; \left.\frac{\partial h}{\partial x}\right|_{x_n} + \frac{A_2}{2} \; \left. \frac{\partial^2 h}{\partial x^2}\right|_{x_n}
\end{eqnarray}
with the parameters
\begin{equation}
A_i = \int_{-\infty}^{\infty} dx \ \delta_N(x-x_n) \cdot (x-x_n)^i .
\end{equation}
The parameter $A_1$ vanishes due to the symmetry of $\delta_N$ and $A_0$ can be calculated as follows,
\begin{eqnarray}
 A_0 &=& \int_{-\infty}^\infty dx \ \delta_N(x) = \int_{-\infty}^\infty \frac{dx}{\sqrt{N}} \ \frac{2}{1+ \left( 1+ \frac{\gamma_n x^2 }{N}\right)^N} \nonumber \\
 &\approx& \frac{1}{\sqrt{N}} \int_{-\infty}^\infty dx \ \frac{2}{1+\exp\left( \gamma_n x^2 \right)} + \frac{\tilde{A_0}}{N^{3/2}} \; ,
\end{eqnarray}
where we applied the identity
\begin{equation}
 \exp(x) = \lim_{N\rightarrow \infty} \left( 1 + \frac{x}{N} \right)^N.
\end{equation} 
The contribution $\tilde{A_0}$ cancels out later and is therefore omitted in the following.
The integral can be expressed in terms of the Riemann zeta function $\zeta(x)$ and we obtain
\begin{equation}
A_0 \approx \frac{\Lambda_0}{\sqrt{\gamma_n N}} \; ,
\end{equation}
where
\begin{equation}
\Lambda_0 = -2(\sqrt{2}-1) \sqrt{\pi} \cdot \zeta(1/2).
\end{equation}
Similarly, we find that
\begin{equation}
 A_2 =  \frac{\Lambda_2}{(\gamma_n N)^{3/2}} \; ,
\end{equation}
where
\begin{equation}
 \Lambda_2 = \frac{(\sqrt{2}-1)\sqrt{\pi}}{\sqrt{2}} \cdot \zeta(3/2).
\end{equation}
Thus, we get the expansion
\begin{equation}
\int dx \langle G \rangle_h (x) h(x) \approx G_0 \ A_0 \left( h(x_n) + \frac{\Lambda}{16} \delta x^2 \left.\frac{\partial^2 h}{\partial x^2}\right|_{x_n} \right) \; ,
\label{eq:deltaE-expansion}
\end{equation}
where 
\begin{equation}
\Lambda = \frac{\Lambda_2}{\Lambda_0} = - \frac{1}{2 \sqrt{2}} \frac{\zeta(3/2)}{\zeta(1/2)} \approx 0.6325
\end{equation}
and the full width at half maximum $\delta x = \sqrt{8/\gamma_n N}$ of the conductance peak of $\langle G \rangle (x)$.

\subsection{Expansion of the thermopower $S_{\langle G \rangle}$}
\label{app:delta_E:thermopower}

Then, we apply the expansion on the integrals the Cutler-Mott equation Eq.~(\ref{eq:cutler_mott}) which we rewrite in the dimensionless variables $x=E/V_0$, $t=T/T_0$ and $\tilde{\mu}=\mu/V_0$,
\begin{equation}
 \SG = -\frac{k_B}{e} \frac{1}{t} \frac{\int dx \ \langle G \rangle_h \left( -\partial_x f \right) (x-\tilde{\mu})}{\int dx \ \langle G \rangle_h \left( -\partial_x f \right)} .
\end{equation} 
We find
 \begin{equation}
 \SG \approx -\frac{k_B}{e} \frac{1}{t}  \left[ \Delta x + \frac{\Lambda}{8} \delta x^2 \frac{\partial}{\partial x} \log \left( -\frac{\partial f}{\partial x} \right)_{x_n} \right] \; ,
\end{equation}
where $\Delta x = x_n - \tilde{\mu}$.
With $\Gamma_n = \gamma_n / V_0^2$ , ${\Delta E = E_n - \mu}$, $\delta x = \delta E / V_0$ and the full width at half maximum ${\delta E \approx \sqrt{8/\Gamma_n N}}$ of the conductance peak we obtain the thermopower in the old notation
\begin{eqnarray}
\SG &\approx& -\frac{1}{e T} \left[ \Delta E + \frac{\Lambda}{8} \delta E^2 \frac{\partial}{\partial E} \log \left( -\frac{\partial f}{\partial E} \right)_{E_n} \right] \nonumber \\
&=& -S_0 \left[ \frac{\Delta E}{k_B T} - \frac{\Lambda}{8}  \frac{\delta E^2}{(k_B T)^2} \tanh\left( \frac{\beta}{2}\Delta E \right) \right] \; , \nonumber \\
\label{eq:app:thermopower}
\end{eqnarray}
where $S_0 = k_B / e$.
We approximate ${\tanh\left( \beta \Delta E / 2 \right) \approx \textrm{sign}(E_n - \mu)}$ such that we eventually end up with the expression
\begin{equation}
\SG(T) \approx - S_0 \left( \frac{|\Delta E|}{k_B T} - \frac{\Lambda}{8} \frac{\delta E^2}{(k_B T)^2}  \right) \cdot \textrm{sign}(\Delta E). \label{eq:thermopower_expansion}
\end{equation}

\subsection{Expansion of the $ZT$ value}
\label{app:delta_E:ZT}

For a derivation of the electrical conductivity and the heat conductivity in the Landauer-B\"uttiker picture we refer to the work of Guttman et al\cite{guttman_thermoelectric_1995}. They are given in terms of the conductance $G$ by
\begin{eqnarray}
 \sigma_{el} &=& K_0 \; ,\\
 \kappa_{el} &=& \frac{1}{e^2 T} \left( K_2 - \frac{K_1^2}{K_0} \right) \; ,
\end{eqnarray}
where 
\begin{equation}
 K_n[G] = \int \ dE \ G(E) \left( - \frac{\partial f}{\partial E} \right) (E-\mu)^n \; .
\end{equation}
Then we can apply the $\delta E$-expansion Eq.~(\ref{eq:deltaE-expansion}) and find after some calculations that
\begin{eqnarray}
\kappa_{el} &=& \frac{A_0 G_0}{e^2 T} \frac{\Lambda}{8} \delta E^2 \left( -\frac{\partial f}{\partial E} \right)_{E_n} \; , \\
\sigma_{el} &=& G_0 A_0 \left( -\frac{\partial f}{\partial E} \right)_{E_n}+ \mathcal{O}(\delta E^2)  \; .
\end{eqnarray}
From Eq.~(\ref{eq:thermopower_expansion}) we know that
\begin{equation}
 S = - \frac{k_B}{e} \frac{\Delta E}{k_B T} + \mathcal{O}(\delta E^2)
\end{equation}
such that the $ZT$ value for $\kappa_{ph}=0$ in the $\delta E$-expansion is given by
\begin{equation}
 ZT = T \frac{\sigma_{el} S^2}{\kappa_{el}} = \frac{8}{\Lambda} \left( \frac{\Delta E}{\delta E} \right)^2 + \mathcal{O}(\delta E^0) \; .
\end{equation}


\bibliography{Thermoelektrika_2.bib}


\end{document}